\documentclass[aps,prb,showpacs,floatfix,amsmath,amssymb,
               superscriptaddress,reprint]{revtex4-1}
\usepackage{color}
\usepackage{graphicx}
\usepackage{natbib}
\usepackage{epsfig}
\usepackage{setspace}
\usepackage{amsmath,amssymb}
\usepackage{verbatim}
\usepackage{bibentry}

\begin{document}
\title{Emergence of non-Fermi liquid dynamics through non-local
correlations in an interacting disordered system}
\author{Sudeshna Sen}\email{sudeshna@sjtu.edu.cn}
\affiliation{Department of Physics and Astronomy, Shanghai Jiao Tong University, Minhang Campus, Shanghai-200240, China}
\author{N. S. Vidhyadhiraja} \email{raja@jncasr.ac.in}
\affiliation{Theoretical Sciences Unit, 
Jawaharlal Nehru Centre for Advanced 
Scientific Research, Bangalore-560064, India}
\author{Mark Jarrell}\email{jarrellphysics@gmail.com}
\affiliation{Department of Physics \& Astronomy, Louisiana State University, Baton Rouge, Louisiana 70803, USA}
\begin{abstract}
\noindent 
We provide strong evidence for a quantum critical point (QCP) associated with the destruction of Kondo screening in the Anderson-Hubbard model for interacting electrons with quenched disorder. The evidence comprises three elements: (a) the identification of an energy scale, $\omega^*$, that
delineates infrared Landau damping from higher frequency
non-Fermi liquid(nFL) dynamics; (b) the finding that this crossover scale $\omega^*$ appears to vanish with increasing disorder; (c) the concomitant appearance of a finite intercept in a broad distribution of Kondo scales. 
Our findings indicate a Kondo destruction scenario, albeit distinct from the local QCP picture. The nFL behavior is shown to stem from an interplay of strong electron-electron interactions and the systematic inclusion of short-range dynamical fluctuations induced by the underlying random potential. The results have been obtained through a computational framework based on the typical medium dynamical cluster approximation. 
\end{abstract}
\maketitle

\section{Introduction}
\label{intro}
The paradigm of Landau's Fermi liquid (FL) theory \cite{Pines_quantum_liquids} provides a robust foundation 
for understanding metals in terms of weakly interacting electron-like `quasiparticles'.  
However, there 
exist several classes of materials where deviations from FL theory have been observed. A universal feature
of such materials is the simultaneous presence of strong electron-electron (e$^-$- e$^-$) repulsive 
interactions and disorder \cite{VLOHNEYSEN1997NFL}. Examples include heavy-fermions \cite{Si_review, 
Steglich_review,Miranda1997_review,NFL_scaling_UPd,NFL_scaling_UCuPd_2,maclaughlin2001glassy,booth1998pd}, rare-earths \cite{rare_earth1, rare_earth2}, cuprates \cite{cuprates1, cuprates2, cuprates3}, and doped semiconductors \cite{kravchenko_review_2004}. Of particular relevance to this 
paper is the breakdown of the FL paradigm in the dual presence of strong e$^-$- e$^-$ interactions ($U$) and 
quenched disorder ($W$)\cite{Miranda1997_review}. 

The origin of non-FL (nFL) behavior in strongly correlated disordered systems has eluded theorists and experimentalists alike \cite{Miranda1997_review} and has thus received a sustained interest \cite{anamitra2017}. 
Some early experimental \cite{bernal1995copper} and theoretical work on the phenomenological Kondo disorder model \cite{vlad_KDM,Miranda1997_review,Jarrell_nFL1997} and 
on microscopic strongly correlated models \cite{MFT_Vlad_Kotliar_1997, Miranda_griffiths_phase2001} showed that responses from anomalously low Kondo scales may be connected to singular thermodynamic responses and nFL behavior. These sites with anomalously low Kondo scales form sparse regions of local moments \cite{Miranda_griffiths_phase2001}, consistent 
with an interpretation in terms of Griffiths effects \cite{Sachdev173,Miranda_griffiths_phase2001}. Furthermore, a 
relatively recent work \cite{Andrade20093167} has highlighted the importance of disorder induced spatial inhomogeneities in such scenarios. 
Experimental imaging of disordered strongly correlated systems 
\cite{Hamidian08112011, HTC_disorder1} reveal the emergent role of disorder induced spatial inhomogeneities on the microscopics of such systems that in turn would influence the thermodynamics. Such experiments 
show how, even non-magnetic randomness in a Kondo system can induce strong hybridization modulation thus influencing the electron scattering dynamics. Proposed mean field approaches neglecting spatial fluctuations due to 
disorder cannot adequately address the specific low energy scales that determine the nFL nature revealed by strongly correlated disordered systems. The theory for such systems should therefore 
comprise two critical ingredients, namely, systematic inclusion of short range dynamical fluctuations due to disorder and its interplay with the local Kondo physics due to strong correlations. 

There now exists compelling evidence of a disorder driven quantum-critical, metal-insulator transition in 
correlated two-dimensional systems \cite{kirkpatrick_review,2016arXiv160306525P} and 
associated nFL charge dynamics \cite{2016arXiv160306525P}. 
The quantum critical nature of the metal-insulator transition in bulk, lattice systems in presence of disorder and Hubbard-type interactions has also been reported \cite{Lee_NiSe,Husmann,NFL_scaling_UPd,NFL_scaling_UCuPd_2,expt_review_Lohneysen}. Theoretically, the quantum critical nature of such a metal-insulator transition in the disordered two-dimensional electron gas was established using the two-loop renormalization group approach\cite{Punnoose2005}. Irrespective of the experimental details, these observations generically support a scenario where at $T=0$ and $W=0$, the system is a normal FL metal gradually developing nFL excitations before undergoing a metal-insulator transition at $W=W_c$, demonstrating critical nFL dynamics in the vicinity of a quantum critical point (QCP).  
Despite the early reports on disorder induced nFL phenomenology based on emergent local moments \cite{Miranda1997_review,MFT_Vlad_Kotliar_1997, Miranda_griffiths_phase2001}, a fundamental challenge remained, namely, 
(1) do these rare regions of `emerging' local moments, dubbed as Griffiths singularities, act as precursors to a `genuine' QCP, and (2) what is the feedback effect of these 
local moment instabilities on the underlying interacting FL from which they emerge? 

\begin{figure}[htp!]
  \centerline{\includegraphics[clip=,scale=0.4]
                          {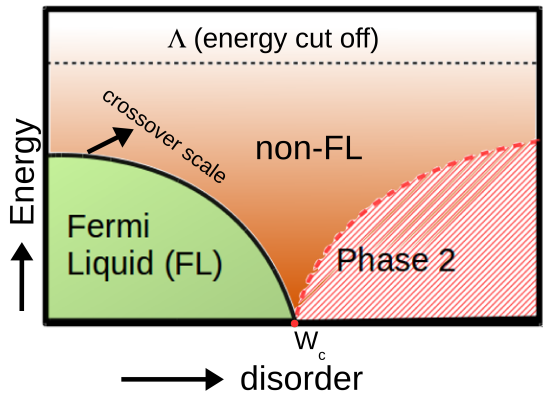}}
  \caption{{\bf Schematic representation of the obtained crossover energy scale separating FL and nFL dynamics:}
  The black solid line represents the crossover scale derived within our $T=0$ calculations. This scale marks a crossover from FL dynamics to nFL dynamics with increasing energy, and its vanishing would eventually lead to a QCP at a critical disorder strength, $W_c$. This dynamics would manifest in the finite temperature fan of the QCP .
The black dotted line represents a high energy cut-off, $\Lambda$, beyond which such a description of the dynamics becomes invalid. The red dashed line separates the nFL phase from a second phase the nature of which cannot be determined within the current theory, but is inferred from previous studies \cite{Lee_NiSe,Husmann,NFL_scaling_UPd,NFL_scaling_UCuPd_2,expt_review_Lohneysen,kirkpatrick_review,2016arXiv160306525P}.
}
  \label{fig:schematic}
\end{figure}

In our work we provide unambiguous evidence of the `existence of such a QCP' starting from the FL phase of a microscopic Hamiltonian, namely the Anderson-Hubbard model (AHM), where the idea of a local moment induced QCP was ruled out in Ref.~\cite{Miranda_griffiths_phase2001}. 
We discern the emergent nFL excitations in proximity to the QCP, by investigating the many body scattering dynamics in the metallic phase.  The schematic presented in Fig.~\ref{fig:schematic} summarizes results from our simulations of 
the AHM and incorporates inferences from previous studies 
\cite{kravchenko_review_2004,expt_review_Lohneysen,QCP_AHM_2D_1,QCP_AHM_2D_2}.  
On the metallic side of the QCP, we find a heretofore unidentified crossover energy scale, $\omega^*$ that appears to vanish at a QCP. 
The system shows characteristic Fermi liquid dynamics for energies, $|\omega|<\omega^*$ and gradually 
deviates from $\sim \omega^2$ dynamics to $|\omega|^\alpha$ for $\omega^*<|\omega|<\Lambda$, as the 
disorder is increased, where the exponent $\alpha$ varies continuously with $W$. For larger $W$'s, beyond the QCP, we assume that there must be some phase transition to quench the entropy associated with the unscreened moments.
 
In this work we establish that the key to quantifying Fig.~\ref{fig:schematic} and understanding the origin of an nFL and a QCP, lies in the systematic incorporation of short range correlations due to disorder into the full many-body scattering dynamics of the electrons. We achieve this by adapting the typical medium dynamical cluster approximation (TMDCA) \cite{Chinedu_TMDCA2014,ekumaSOPT} such that the physics due to multiple scales could be handled. Within this framework, (1) we explore the precise evolution of the disorder averaged scattering dynamics and the associated distribution of Kondo scales; (2) subsequently, we predict the emergence of a disorder-induced nFL dynamics and a unique disorder dependent FL to nFL crossover scale that presumably leads to the QCP, thus delineating the FL-nFL boundary in Fig.~\ref{fig:schematic}. We emphasize that an 
identification of `Phase 2' is beyond the scope of the current formalism and that 
we can only probe the disorder-driven transition starting from a Fermi liquid phase.

The paper is organized as follows. We discuss the Model and Theoretical framework in Section~\ref{framework}, followed by Results and Discussions in Section~\ref{results}. 
We finally conclude in Section~\ref{conclusions}.

\section{Model and Theoretical Framework}
\label{framework}
We investigate the Anderson-Hubbard model (AHM)
for describing the physics due to the interplay of disorder and electron-electron interactions, 
\begin{align}
  \mathcal{H}=\sum_{ij, \sigma}t_{ij}c_{i\sigma}^{\dagger}c^{\phantom\dagger}_{j\sigma}+
              \sum_{i, \sigma}(V_i-\mu)\hat{n}_{i\sigma}+
              U\sum_i\hat{n}_{i\uparrow}\hat{n}_{i\downarrow}
  \label{eq:AHM}
\end{align}
where, $c_{i\sigma}^{\dagger}$ ($c_{i\sigma}$) is the fermionic creation (annihilation) operator for an electron with spin $\sigma$ at site $i$, and $\hat{n}_{i\sigma}=c_{i\sigma}^{\dagger}c^{\phantom\dagger}_{i\sigma}$; $t_{ij}$ is the nearest neighbor hopping amplitude, $U$ is the onsite Coulomb interaction energy. The lattice is represented by a 3D cubic density of states (DoS) with full bandwidth, $D=3$ eV. The random disorder potential, $V_i$, is drawn from a box distribution $P(V_i)$ of width $W$ and represented as $P(V_i)=\frac{1}{2W}\Theta(W-|V_i|)$, where $\Theta(x)$ is a step function.  The disorder averaging is represented using the shorthand notation, $\langle\hdots\rangle=\int dV_i P(V_i)(\hdots)$. The particle-hole (p-h) symmetry is imposed by setting $\mu=U/2$. We define the onsite energy as, $\epsilon_i=-U/2+V_i$ for the rest of the paper.

\begin{figure*}[htp!]
\centerline{\includegraphics[clip=,scale=0.4]
                        {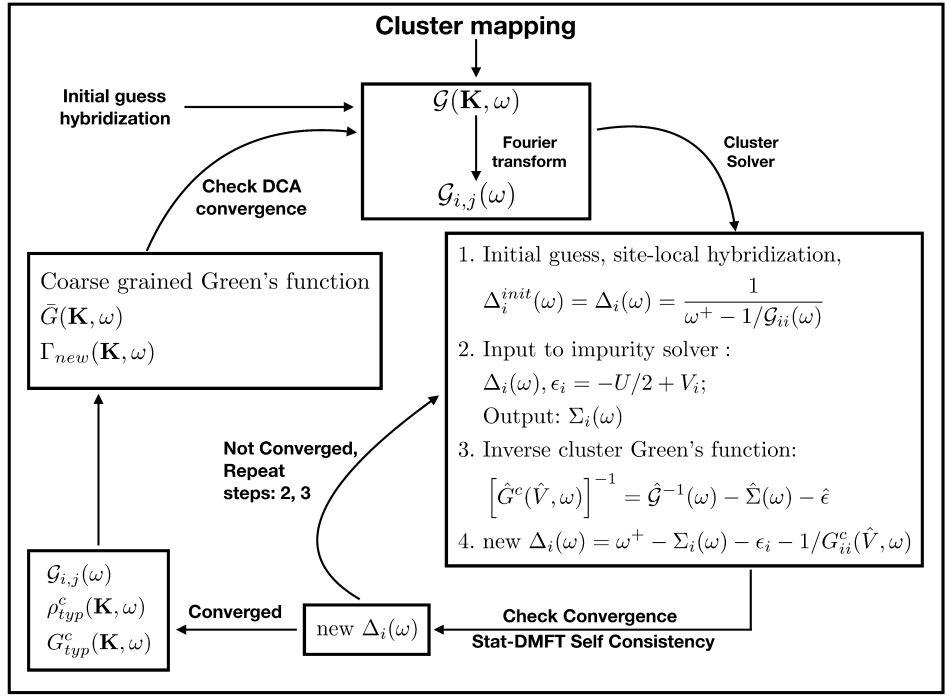}}
  \caption{{\bf Flowchart representing the multiscale approach used in the current calculations.} The TMDCA   self-consistency ensures the systematic and explicit incorporation of short-range correlations, due to disorder into the (stat-DMFT like) loop that utilizes the local moment approach to solve for the strong correlation problem at a single site level.}
  \label{fig:flowchart}
\end{figure*}

The AHM has been explored using various methods like quantum Monte Carlo \cite{QMC_AHM_1,QCP_AHM_2D_2}, dynamical mean field theory based approaches \cite{Vollhardt_review, Vollhardt_DMFT_AHM,Aguiar_2006,Aguiar2009,aguiar_2003,aguiar_2013,Byczuk2005}, and Hatree-Fock based approaches \cite{Sachdev173,Logan1993}. However, in order to understand the dynamical signatures of the QCP,  more sophisticated and advanced theory is required. The theory should be able to tackle the dynamical scales generated by strong correlations and its interplay with the spatial fluctuations brought on by disorder; hence we require a multiscale approach. 
We develop a multiscale approach where we incorporate the {\it dynamical spatial fluctuations} due to disorder within the framework of TMDCA \cite{Chinedu_TMDCA2014}. 

The TMDCA \cite{Chinedu_TMDCA2014} is based on the same self-consistent framework of the standard DCA \cite{hettler1998nonlocal, maier2005quantum,Jarrell_HRK_DCA_disorder}.  However, the crucial difference with the standard DCA lies in the utilization of an appropriately disorder averaged, (momentum, $\mathbf{K}$) dependent hybridization, $\Gamma(\mathbf{K},\omega)$. One starts with the usual DCA cluster mapping of a $d$-dimensional periodic (or disorder averaged to restore translational invariance) lattice 
in momentum space. The cluster consists of $N_c=L_c^d$ cells in $d$ dimensions, with $\mathbf{K}$ being the cell momentum and $L_c$ being the linear dimension of the cluster. This cluster is then embedded into a self consistently obtained effective medium, given by $\Gamma(\mathbf{K},\omega)$.  We now outline the steps below:
\begin{enumerate}
\item{While initializing the problem, one can consider it to be a uniform field, given by $\Gamma_{init}$.}
\item{With this, one can obtain the cluster excluded Green's function, $\mathcal{G}(\mathbf{K},\omega)$, given by, 
$\mathcal{G}(\mathbf{K},\omega)=\left[\omega^{+}-
               \Gamma_{init}-\bar{\epsilon}_\mathbf{K}\right]^{-1}$, 
where $\bar{\epsilon}_\mathbf{K}$ is the coarse-grained bare dispersion.  Hence, spatial correlations up to a range $\xi\lesssim L_c$ are explicitly retained, while the longer length scale physics are described at a mean-field level.}
\item{$\mathcal{G}(\mathbf{K},\omega)$ is then Fourier transformed to get the real space cluster excluded Green's function, $\mathcal{G}_{i,j}(\omega)=\sum_\mathbf{K}\;\mathcal{G}(\mathbf{K},\omega)
\exp[i\mathbf{K}.(\mathbf{r}_i-\mathbf{r}_j)]$}
\item{Then for a given disorder configuration, $\hat{V}$, we may calculate the cluster Green's function $\hat{G}^{c}(\hat{V}, \omega)$ with the effects of disorder and electron-electron interactions incorporated.}
\end{enumerate}
 \begin{figure*}[tp!]
  \centerline{\includegraphics[clip=,scale=0.3]
                        {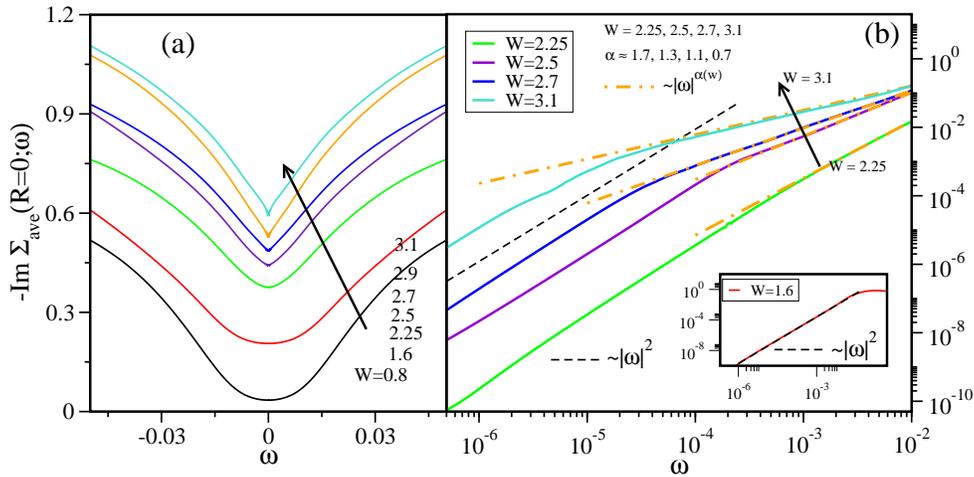}}
   \caption{{\bf Disorder averaged electronic self energy:} 
    (a) The low energy part of the average self-energy, -Im$\Sigma_{ave}(\mathbf{R}=0;\omega)$ is shown for $U=1.6,\,N_c=38$ and various $W$'s as indicated. A clear crossover from a
    Fermi liquid, $\sim \omega^2$ to non-Fermi liquid $\sim |\omega|^\alpha$ is observed. On a linear scale this features as the formation of a cusp at low energies. 
    (b) (Main panel) Some representative data 
    ($W=2.25, \; 2.5, \; 2.7, \; 3.1$) of (a) are plotted on a log-log scale to deduce the power, $\alpha$ as a function of $W$ and also deduce the crossover frequency, $\omega^*$ (using a procedure demonstrated in Figure~\ref{fig:SE_analysis} of Appendix~\ref{appendix0}), beyond which -Im$\Sigma_{ave}(\mathbf{R}=0;\omega)$ evolves from $\sim \omega^2$ (black dashed line) to
    $|\omega|^{\alpha(W)}$ (orange dashed-dotted line), where $\alpha(W)\approx1.7,\; 1.3, \;1.1, \; 0.7$ for 
    $W=2.25, \; 2.5, \; 2.7, \; 3.1$ respectively, as shown in main panel. Note that a value approximately equal to -Im$\Sigma_{ave}(\mathbf{R}=0;\omega=0)(=a_0)$ has been subtracted.
    (b) (Inset) A similar analysis as in (b) (main panel) is illustrated for a relatively lower $W$, namely, $W=1.6$. 
   Here, the crossover occurs at a sufficiently higher energy identified as the peak of the respective distribution (shown in Fig.~\ref{fig:low_disorder_TK}) of Appendix~\ref{appendix3}.}
      \label{fig:SE}
 \end{figure*}
We now discuss the second stage of the self-consistent computational setup involving the treatment of the electron-electron interactions.  Our primary focus is to explore the influence of disorder induced dynamical spatial fluctuations on the Kondo physics governed by $U$. This requires us to use a non-perturbative many-body impurity solver that can capture the single particle spectrum over all energy scales efficiently. 
We utilize the local moment approach (LMA) \cite{LMA_SIAM1,LMA_asymm_SIAM,LMA_NRG_benchmark1,Eastwood1998} in order to obtain the interaction self-energy, $\hat{\Sigma}(\omega)$ that is calculated in real space. Each site in the TMDCA cluster is mapped on to a single impurity Anderson model, the self-energy of which is calculated using the LMA.  We now outline the steps involved in this self-consistency below: 
\begin{enumerate}
\item{For this part of the calculation, we utilize the diagonal part of  $\mathcal{G}_{i,j}(\omega)$ calculated previously. This serves as an initial guess input site-local hybridization, $\Delta_i^{init}(\omega)=\frac{1}{\omega^+-1/\mathcal{G}_{ii}(\omega)}$ to the impurity 
solver along with the local site energy, $\epsilon_i$.}
\item{The inverse cluster Green's function, $ \left[\hat{G}^{c}(\hat{V}, \omega)\right]^{-1}=   \hat{\mathcal{G}}^{-1}(\omega)-\hat{\Sigma}(\omega)-\hat{\epsilon}$ is then calculated utilizing the interaction self-energy, $\hat\Sigma(\omega)$ obtained from the impurity solver.  Note that, $\hat{\Sigma}(\omega)$ and $\hat{\epsilon}$ are diagonal matrices and $\hat{\mathcal{G}}(\omega)$ has off-diagonal elements.}
\item{A new $\Delta_i(\omega)$ is calculated using the relation, $\Delta_i(\omega)=\omega^+-\Sigma_i(\omega)-\epsilon_i- 1/G^c_{ii}(\hat{V},\omega)$;}
\item{in the subsequent iterations within the cluster solver, each site is thus provided with a {\it site dependent} $\Delta_i(\omega)$ input to the impurity solver.}
\item{The iterative loop within the real-space cluster solver is repeated until $\int \Delta_i(\omega)d\omega$ converges for all $N_c$ sites within some tolerance.} 
\end{enumerate}
Note that this scheme resembles the stat-DMFT formulation in the sense that the diagonal Green's functions of a finite $(N_c\times N_c)$ real-space cluster are being solved self-consistently within a DMFT like scheme. In all the results presented below, we use $N_c=38$ and $U=1.6$ unless otherwise specified.

The converged $\hat{G}^{c}(V, \omega)$, from the above cluster solver in 
real space, is Fourier transformed to $\mathbf{K}$ space and the typical density of states, 
$\rho^c_{typ}(\mathbf{K}, \omega)$ is constructed using the following ansatz:
$  \rho^c_{typ}(\mathbf{K}, \omega)=\exp\left(
  \frac{1}{N_c}\sum_{i=1}^{N_c}
  \left\langle\ln\; \rho_i^c(\omega, \hat{V})
\right\rangle
  \right)
  \left\langle\frac{\rho^c(\mathbf{K},\omega,\hat{V})}
  {\frac{1}{N_c}
\sum_i\rho_i^c(\omega,\hat{V})}\right\rangle$. 
The typical cluster Green's function, $G^c_{typ}(\mathbf{K},\omega)$ is then obtained via Hilbert transform of $\allowbreak G_{typ}(\mathbf{K},\omega)=\int\frac{\rho_{typ}
(\mathbf{K},\omega^\prime)d\omega^\prime}{\omega-\omega^\prime}$. 
The coarse-grained Green's function, $\bar{G}(\mathbf{K},\omega)$ is then calculated via, $\allowbreak\bar{G}(\mathbf{K},\omega)=\int\; \frac{N_0^c(\mathbf{K},\epsilon)\;d\epsilon}{\left[G^c_{typ}(\mathbf{K},\omega)\right]^{-1}+\Gamma(\mathbf{K},\omega)-\epsilon+\bar{\epsilon}(\mathbf{K})}$, where, $N_0^c(\mathbf{K},\epsilon)$
represents the bare partial DoS with which we can further calculate the 
new momentum dependent hybridization, $\Gamma(\mathbf{K},\omega)$ as 
$\Gamma_{new}(\mathbf{K},\omega)=
\Gamma_{old}+\zeta\left[(G^c_{typ})^{-1}-(\bar{G})^{-1}\right]$, where 
$\zeta$ is a mixing factor used to get smooth convergence and is typically set 
to a value of $0.5$. At convergence, $G^c_{typ}(\omega)=\bar{G}(\omega)$ within 
some tolerance. We outline this whole procedure in the form of a flowchart in Fig.~\ref{fig:flowchart}.

We now conclude this section with a brief discussion about the LMA. 
This technique has been successfully utilized in several impurity~\cite{LMA_SIAM1,LMA_GAIM,LMA_NRG_soft_gap_AIM,LMA_pseudogap_AIM1} and lattice models within DMFT~\cite{LMA_KIs,Raja_dyn_sca_pam,Raja2_2005,Eastwood1997,hbar}, describing both FL and nFL phases of the respective models. 
In fact, the LMA, although approximate, has been shown to capture almost all of the aspects of Kondo physics in the conventional single impurity Anderson model, in ``an almost exact" way, as evident from the agreement with Bethe Ansatz, numerical renormalization group~\cite{LMA_NRG_benchmark1,Dickens2001,field_dep_BA_LMA,T_BA_LMA}, and even with experiments\cite{LMA_KIs,Raja2_2005}. 
Moreover, the LMA has been successfully utilized (and benchmarked with numerical renormalization group calculations) to understand the physics due to {\it local} quantum phase transitions between a Fermi liquid (Kondo screened phase) and a local moment phase consisting of unquenched `impurity' moments~\cite{LMA_pseudogap_AIM1,LMA_GAIM,LMA_NRG_soft_gap_AIM}. 
Recently, it has also been applied to disordered systems within coherent potential approximation~\cite{CPA_NFL,sen2015spectral}, and again, good agreement with NRG results was found, and a new non-Fermi liquid mechanism was also proposed. 
LMA was also implemented for the Anderson-Hubbard model within typical medium theory~\cite{my_paper_TMT}, which was again in good agreement with NRG calculations~\cite{Byczuk2005}.

The current implementation of LMA considers infinite resummation of a specific class of diagrams describing dynamical spin-flip scattering processes inherent to the physics of the Kondo effect. 
We refer interested readers to several previous works for more details, e.g.~\cite{LMA_SIAM1,LMA_GAIM,LMA_pseudogap_AIM1,LMA_KIs,Raja_dyn_sca_pam}.
Thus, the current formalism of LMA makes it a perfectly suitable tool for handling local quantum critical points involving criticality due to the breakdown of Kondo screening.
These spin-flip processes generate a peak at a low energy scale, $\omega_m$, in the imaginary part of the transverse spin polarization propagator, where $\omega_m$ is of the order of the Kondo scale, $T_K$. 
In clean systems the Fermi liquid to local moment formation is therefore signaled as a spin-flip pole in the imaginary part of the transverse spin polarization propagator. 
In a disordered system this pole would occur at certain sites where the local moment forms.

The generalization of LMA to symmetry breaking phase transitions and cluster geometries involving non-local Coulomb interaction effects is not straightforward.
The many-body diagrams considered within the LMA should be accordingly adapted to handle such situations.
The extension of LMA to clusters has been attempted in the present work, in an approximate way, through an integration of the DCA (and TMDCA) with a stat-DMFT like cluster solver based on LMA.
This extension for a disordered interacting system, indeed deciphers a generic microscopic mechanism for observing non-Fermi liquid dynamics due to disorder and local Coulomb interactions as described in the subsequent sections of this paper. However, we agree that a true cluster extension of LMA has not yet been carried out. 

\section{Results and Discussions}
\label{results}
\subsection{Scattering dynamics}
One of the main highlights of this work is the identification of a 
{\it critical} low energy scale, $\omega^*$ such that in presence of 
disorder the scattering dynamics has the usual FL form only at energies, 
$\omega<\omega^*$. In order to identify $\omega^*$ we probe the imaginary 
part of the disorder averaged electronic self-energy, $\Sigma_{ave}(\mathbf{K},\omega)$, 
obtained from the 
Dyson's equation involving the {\it arithmetic} 
average of $G^c(\mathbf{K},\omega)$, (the average 
being denoted as $\langle G^c(\mathbf{K},\omega)\rangle_{ave}$),
that in turn is obtained from the Hilbert 
transform of $\langle\rho^c(\mathbf{K},\omega,V)\rangle$, where, 
$\rho^c(\mathbf{K},\omega,V)=-\frac{1}{\pi}\mathrm{Im}G^c(\mathbf{K},\omega,V)$. 
The disorder 
averaged self-energy, $\Sigma_{ave}(\mathbf{K},\omega)$ is then obtained as following,
\begin{align}
\Sigma_{ave}(\mathbf{K},\omega)=\mathcal{G}^{-1}(\mathbf{K},\omega)-
\langle G^c(\mathbf{K},\omega)\rangle_{ave}^{-1},
\label{eq:SE}
\end{align}
with the local self-energy being, $\Sigma_{ave}(\mathbf{R}=0,\omega)=\sum_{\mathbf{K}}\Sigma_{ave}(\mathbf{K},\omega)$.

In Figure~\ref{fig:SE}(a) we plot $-\mathrm{Im}\Sigma_{ave}(\mathbf{R=0};\omega)$ as a function of increasing $W$, with frequency plotted on a linear scale. 
The non-zero, $\omega=0$, contribution ($a_0$) in the self-energy represents the static elastic impurity scattering, while the $|\omega|\rightarrow 0$ has both {\it inelastic} and elastic contributions. 
Also, physically consistent is the observation that at sufficiently weak disorder, {\it e.g.} $W=0.8$, $a_0$ is sufficiently small such that we can expect a Drude like expression and an arbitrarily large d.c.\ conductivity. 
This picture however breaks down as one increases $W$. 
Beyond a certain disorder strength, $W=2.25$, {\it even on a linear scale}, the lineshape develops a clear cusp in the immediate vicinity of the Fermi energy. 
Thus, for $W\gtrsim 2.25$, we can identify a vanishingly small emergent low energy scale, $\omega^*$ beyond which the FL behavior crosses over to nFL dynamics. 
Thus, for $\omega^*<|\omega|<\Lambda$, the scattering dynamics can be represented by a power law energy dependence, given by, $-\mathrm{Im}\Sigma_{ave}(\mathbf{R}=0;\omega)\sim|\omega|^{\alpha(W)}$, where $\Lambda$ is high energy cut-off frequency. 
This is further highlighted in Figure~\ref{fig:SE}(b) where we plot the low frequency regime of $-\mathrm{Im}\Sigma_{ave}(\omega)-a_0$ on a log-log scale. 
The $\sim |\omega|^{\alpha(W)}$ functional dependence is shown as the orange dashed-dotted lines in Figure~\ref{fig:SE}(b) and the exponent, $\alpha$, obtained with this fitting is found to be dependent on $W$.

\begin{figure}[ht!]
  \centerline{\includegraphics[clip=,scale=0.5]
                          {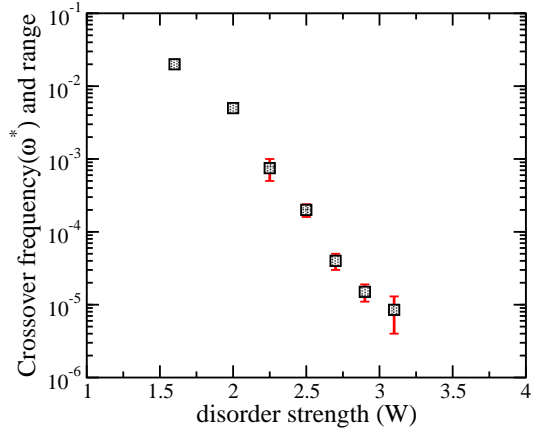}}
  \caption{{\bf FL to nFL crossover energy scale ($\omega^*$) and energy range ($\delta\omega^*$):} 
  The $\omega^*$ (shaded squares) and $\delta\omega^*$ (vertical bars) are estimated using the method described in the caption of Figure~\ref{fig:SE_analysis} and are plotted as a function of $W$. Both $\omega^*$ and $\delta\omega^*$ decrease rapidly as $W$ increases providing evidence of the approach towards a disorder driven quantum critical point that separates a Fermi liquid phase from a non-Fermi liquid phase at higher disorder.
  This quantifies the boundary marked as `crossover scale' in Fig.~\ref{fig:schematic}.}
  \label{fig:nFL_scale}
\end{figure}

Furthermore, a closer look at the data reveals a region of  frequencies over which the Im$\Sigma_{ave}$ crosses over from $\omega^2$ to $|\omega|^{\alpha(W)}$ dynamics.  
Thus, for each disorder strength we should not only estimate an approximate crossover point, $\omega^*$ but also a crossover region, $\delta\omega^*$.  
An example of such an analysis is shown in Figure~\ref{fig:SE_analysis}(a,b) of Appendix~\ref{appendix0}. 
In Figure~\ref{fig:nFL_scale} we therefore plot the extracted $\omega^*$ as a function of $W$ (represented as shaded squares) and also mark the crossover region, $\delta\omega^*$, as vertical bars. 
We observe that with increasing disorder, both $\omega^*$ and $\delta\omega^*$ decrease sharply providing evidence for an ensuing QCP. 
The crossover frequency, $\omega^*$, thus  emerges as a {\it unique} vanishing energy scale indicating the emergence of a disorder induced non-Fermi liquid at a critical $W=W_{c}$. 
In other words, the emergence of an arbitrarily small FL to nFL crossover energy scale, $\omega^*$, with increasing disorder, should lead to a qualitative change in the ground state at $W=W_c$.   
As summarized in the schematic represented by Fig.~\ref{fig:schematic}, our calculations thus bring out the (low-) energy boundary that separates the non-Fermi liquid physics of the disorder driven QCP from the conventional FL. 

Finally, we note that the $\omega^*$ in Figure~\ref{fig:nFL_scale} appears to deviate and even slightly approach saturation. 
We speculate that this is due to numerical intractability of the system close to the localization transition, and that the crossover scale should in fact vanish eventually at a finite critical disorder, $W_c$. 
The main reasons for such a numerical bottleneck are that:
(i) the regime of vanishing $\omega^*$ entails impurity sites with vanishing Kondo scales, which is very hard to capture, and 
(ii) the Anderson-Mott insulating regime exhibits a spectral function with singularities that appear as a line of poles, which is again very difficult to capture numerically. 
The apparent saturation is thus a result of this numerical drawback. 
The physical reasons behind this numerical difficulty also indicate that one needs to go beyond the current formalism to be able to simultaneously handle the local moments that would form in presence of the remaining Kondo screened moments. 

Furthermore, the existence of a finite critical disorder has been found in three works, which have considered either a closely related system or the same system as in the present manuscript. 
(a)The 2-loop RG work of Finkelstein and Punnoose~\cite{Punnoose2005} established the existence of a QCP at a finite $W_c$ in an interacting, disordered two dimensional electron gas with an anomalous enhancement in the magnetic susceptibility near the QCP.
(b) Typical medium theory calculations ($N_c=1$ limit of TMDCA) using NRG
as the impurity solver~\cite{Byczuk2005}, and using LMA by the present authors~\cite{my_paper_TMT}, have established that the critical disorder is finite in the disordered Hubbard model. 
(c) Recent exact TMDCA calculations (by two of our current authors)~\cite{ekumaSOPT} indicate a finite critical disorder for Anderson localization, $W_c^{AL}$  in the weak-coupling regime of the three-dimensional Hubbard model. 
The realization of a vanishingly small $\omega^*$ indicates emergent local moment formation that is naturally influenced by the states at the band center because of Kondo effect. 
Furthermore, these states have a natural tendency to Anderson localize at $W_c^{AL}$. 
Thus, although $W_c$, at which $\omega^*$ vanishes, may not coincide with $W_c^{AL}$, it can definitely be considered as a lower bound for Anderson localization of the system, i.e $W_c\leq W_c^{AL}$. Thus, if the latter is finite, the former must also be finite. 

\begin{figure}[htp!]
\centerline{\includegraphics[clip=,scale=0.5]
                        {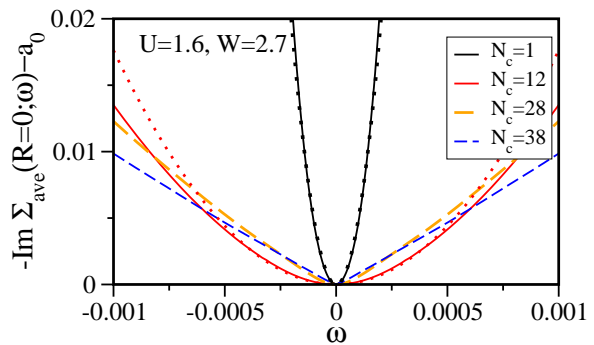}}
  \caption
{{\bf Comparison of the low energy dynamics of the imaginary 
part of the average self-energy for different cluster sizes:}
-Im$\Sigma_{ave}(\mathbf{R}=0,\omega)$ is shown for $U=1.6$ and
$W=2.7$ with $a_0=$-Im$\Sigma_{ave}(\mathbf{R}=0,\omega=0)$ subtracted.
The dotted lines represent fits to the Fermi liquid form, given
by $\sim a\omega^2$. While small cluster sizes of $N_c=1,\,12$
fit well to this form, the systematic deviation is clearly
noticeable for $N_c=28,\,38$ evident from the development of a low energy cusp. Moreover, for both $N_c=28$ and $N_c=38$, the Im$\Sigma_{ave}(\mathbf{R}=0,\omega)$ demonstrate very similar frequency dependence, with estimated $\alpha(W)\approx 1.2,\;1.1$ for $N_c=28,\;38$ respectively, indicating a rapid saturation of $\alpha(W)$ with increasing $N_c$.}
  \label{fig:Nc_compare}
\end{figure}

The systematic incorporation of the short range 
spatial fluctuations due to disorder is of paramount importance in order to observe such nFL scattering. 
We corroborate this through Figure~\ref{fig:Nc_compare} where we plot the imaginary part of the average local self-energy, $\Sigma_{ave}(\mathbf{R}=0;\omega)$, with $a_0$ subtracted, for different cluster sizes, namely, $N_c=1,\,12,\,28,\,38$. 
We emphasize that, the analysis of the low energy frequency dependence of the scattering dynamics is better understood with the subtraction of $a_0$. (For a comparison of $a_0$ we urge the reader to refer to Figure~\ref{fig:SE_a0_Nc} in Appendix~\ref{appendix1}.)  
This comparison illustrated in Figure~\ref{fig:Nc_compare} highlights that the systematic inclusion of short range correlations due to disorder with increasing $N_c$ is absolutely crucial for the non-Fermi liquid dynamics manifested through the  vanishingly small, emergent energy scale $\omega^*$. 
Clearly, the low energy frequency dependence of -Im$\Sigma_{ave}(\mathbf{R}=0,\omega)$ for $N_c=28$ and $N_c=38$ are hard to distinguish, with $\alpha\approx1.2,\;1.1$ for $N_c=28,\;38$ respectively. (A similar tendency is noted for $a_0$ as the $N_c$ is systematically increased as illustrated in Figure~\ref{fig:SE_a0_Nc} in Appendix~\ref{appendix1}.)
The respective $\omega^*\approx10^{-4},\;4\times10^{-5}$ for $N_c=28,\;38$ respectively.
This would thereby imply a strong tendency for $\alpha(W)$ to saturate with increasing $N_c$ and additionally suggests that $N_c=38$ is indeed close to the true thermodynamic limit. 

While a precise statement about the absolute value of $\alpha(W)$ should involve analysis for $N_c>38$, the evidence of an $\alpha(W)$ considerably less than the Fermi liquid value of $2$ is already guaranteed by $N_c=28$ and $N_c=38$.
The rapid approach to the thermodynamic limit within the TMDCA is encouraging and is also in agreement with recent calculations on the 3D Anderson disorder model for non-interacting~\cite{Chinedu_TMDCA2014} and weakly interacting systems~\cite{ekumaSOPT} with the TMDCA framework, that have also shown rapid convergence as a function of increasing cluster size. 

 Additionally, as a manifestation of the nFL scattering dynamics, the spectral lineshape, shown in Appendix~\ref{appendix2}, also develops a cusp, in the vicinity of the Fermi-energy that becomes more pronounced as $\omega^*\rightarrow 0$ and one approaches QCP. 
 The evolution of the low energy spectral lineshape is depicted in Fig.~\ref{ADOS1} of Appendix~\ref{appendix2}. 
 In Figure~\ref{ADOS2} of Appendix~\ref{appendix2} we also report the behavior of the {\it typical} density of states, $\rho_{typ}(\omega)$ and the {\it arithmetically averaged} density of states, $\rho_{arith}(\omega)$ over all energy scales for different $W$'s and two different interaction strengths, namely $U=1.6,\;2.0$.

Finally, how would this observation manifest in the temperature dependence of the resistivity?
Let us assume that there are no vertex corrections just like in infinite dimensions. Then the DC resistivity is given by,
$\sigma_{DC}=\int_{-\infty}^{\infty}d\omega \left(-\frac{\partial n_F}{\partial \omega}\right)\frac{1}{2|\mathrm{Im}\Sigma(\omega)|}$.
At zero temperature, $T=0$, this corresponds to a finite resistivity determined by the elastic scattering off the random potential. It can be readily seen that at $T\neq0$, in a clean FL, where $\mathrm{Im}\Sigma(\omega)\sim \omega^2$ this corresponds to $\sim T^2$ behavior of the resistivity at low temperatures, when $T\ll T^*$ where $T^*$ is the FL coherence scale represented by the Kondo scale in infinite dimensions. Let us now apply this naive picture to the present calculations, 
where we may express $\mathrm{Im}\Sigma(\omega\to 0)$ as, $\mathrm{Im}\Sigma(\omega\to 0)\sim a_0+\omega^2\Theta\left(|\omega^*|-|\omega|\right)+|\omega|^\alpha\Theta\left(|\omega^{**}|-|\omega|\right)\Theta\left(|\omega|-|\omega^*|\right)$. 
Clearly, now $\rho(T)$ crosses over from a $T^2$ behavior to a 
$T^{\alpha(W)}$ behavior thus bearing signatures of a nFL beyond a temperature, $T^*$ associated with $\omega^*$. While such an $\mathrm{Im}\Sigma$ has indeed been observed in this work, the above expression should be used with care. This is because the vertex corrections have been completely ignored even though we are treating short-range fluctuations due to the random potential. Nevertheless, it paves the way for a manifestation of the nFL scattering dynamics on the transport quantities in such systems, and 
the complete analysis including vertex corrections is left as a future challenge. 

\begin{figure}[htp!]
  \centerline{\includegraphics[clip=,scale=0.5]
                        {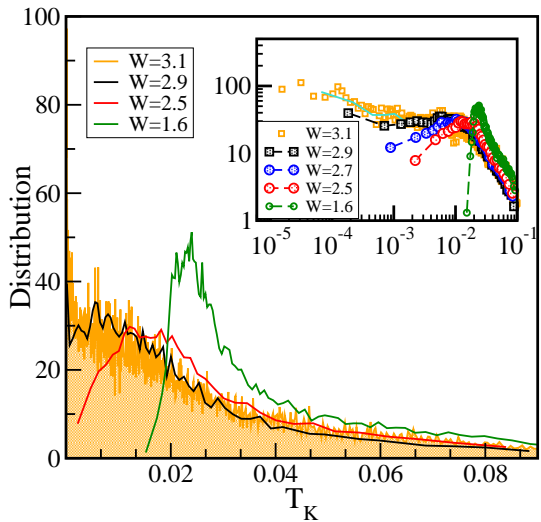}}
   \caption{{\bf{Distribution of Kondo scales:}} In presence of disorder, one obtains a distribution of Kondo scales as shown    in the main panel on a linear scale for $U=1.6$, $N_c=38$ and increasing $W=1.6,\,2.5,\,2.9,\,3.1$.  For low enough disorder, $W=1.6$, the distribution is marked by the presence of prominent maximum and a lower $T_K$ tail bounded from below. At intermediate $W$, ($W=2.5$), even lower scales emerge indicated by a broader lower $T_K$ tail and a broader maximum of the distribution that gradually merges with the tail. At sufficiently, large $W$, ($W=3.1$), as $\omega^*\to 0$ (in Fig.~\ref{fig:nFL_scale}), the distribution tends to acquire a finite intercept $(P(T_K=0)$ as $T_K\rightarrow 0$. This becomes evident in the inset. (Inset) $P(T_K)$ is plotted on a log-log scale to highlight the evolution of the lower $T_K$ tail.}   
  \label{fig:Tk_dist}
\end{figure}

\subsection{Distribution of Kondo scales}
\label{PTk}
We measure the distribution of Kondo scales $P(T_K)$ on the cluster following the method we employed previously for $N_c=1$~\cite{my_paper_TMT}.  
In our previous study for $N_c=1$ \cite{my_paper_TMT}, we found that $P(T_K)$ was dominated by a single sharp peak defining a typical value at the bottom of the distribution.  As a result, the calculations reveal a FL characterized by this typical value. In the current calculations we find that as we increase the cluster size $N_c$ the distribution of impurity environments also increases yielding a broader distribution $P(T_K)$. 
(In Fig.~\ref{fig:Nc_compare} of Appendix~\ref{appendix1} we also plot the self energy for different $N_c$'s and see that the nFL character emerges with increasing $N_c$.)  We now look into the evolution of $P(T_K)$ by gradually increasing $W$, at a fixed $U=1.6$ and cluster size, $N_c=38$. The main panel of Fig.~\ref{fig:Tk_dist} illustrates $P(T_K)$ on a linear scale.  At lower $W$'s, $P(T_K)$ demonstrates a prominent peak, $T_K^{peak}$ leading to largely Kondo Fermi liquid formation at roughly this energy scale. As we increase $W$, the lower $T_K$ tails grow further spanning even lower energies, such that at $W=2.9$ the tail merges with the $T_K^{peak}$ resulting in a broad distribution.
At such $W$'s, extremely low Kondo scales i.e $T_K\lesssim 10^{-5}$, emerge to be highly probable. For higher $W$'s (e.g. $W=3.1$), as $\omega^*\to 0$, the $P(T_K)$ tends to acquire a finite intercept $(P(T_K=0)$ as $T_K\rightarrow 0$. This behavior is even more evident in the inset of Fig.~\ref{fig:Tk_dist} where the respective $P(T_K)$ is plotted on a log-log scale. Fig.~\ref{fig:Tk_dist} is in fact reminiscent of the $P(T_K)\sim T_K^{\beta-1}$ form obtained in earlier calculations in the strong coupling limit (\cite{MFT_Vlad_Kotliar_1997,Vlad_griffiths_phase, aguiar_2013}, that related the $W$'s with $\beta<1$ to electronic Griffiths phases and an associated nFL behavior of the response functions. 

Strongly correlated systems with prevalent Kondo screening currently offer two kinds of QCP\cite{Steglich_review, Si_review}; one represents the conventional Hertz-Millis-Moriya scenario in which the Kondo scale remains non-vanishing even after a magnetic transition. In the other kind, dubbed the local quantum critical scenario, the magnetic transition is accompanied by the breakdown of the Kondo singlet. Our calculations show that the vanishing low energy scale, $\omega^*$ identifying the QCP is concomitant with $P(T_K=0)$ being non-zero. Thus we may infer a Kondo destruction scenario of the associated QCP, driving the system towards formation of local moments, an essential aspect of strongly correlated disordered systems \cite{Paalanen1988,Milov1989,Bhatt1,Bhatt1992,Sachdev_Bhatt}.  However, the simultaneous occurrence of a broad $P(T_K)$, spanning several orders of energy scales also emphasize a mechanism different from that of the local QCP picture \cite{Si_Steglich_QCP_review}. The spatial inhomogeneity due to disorder plays an essential role in causing Kondo destruction at a finite fraction of sites, while the remaining sites
continue to retain local FL character across the $\omega^*\rightarrow 0$ transition.  Our results thus demonstrate a different kind of QCP. It differs from the local QCP in the exponents in the QC regime, e.g.\ $\alpha$, changes continuously as the QCP is approached. With increasing disorder, presumably, there would be a `Phase 2', as illustrated in Fig.~\ref{fig:schematic}, identifying the nature of which is beyond the scope of the current work.  It is however worth mentioning that self-consistent unrestricted Hartree Fock studies \cite{Logan_AHM1, Logan_AHM2, Logan1993, Sachdev173} predict distinct mean-field  magnetic ground states owing to the formation of local moments including a spin glass phase \cite{Logan1993, Sachdev173, GA_spinglass,dobrosavljevic2012conductor}. 

\section{Conclusions}
\label{conclusions}
 We present a study 
investigating the influence of short-ranged, dynamical fluctuations due to disorder on the effective Kondo screening in a disordered, strongly correlated system, within the TMDCA framework. We focus on the disorder averaged scattering dynamics. Our findings reveal the existence of an intrinsic energy scale, $\omega^*$, that behaves like a critical boundary separating the disorder induced nFL dynamics from the conventional FL scattering in strongly correlated disordered systems. In other words as the local moments emerge, exemplified by the conventional phenomenology of a broad distribution of Kondo scales, 
an intrinsic energy scale of the global system, namely, $\omega^*$ also tends to continuously vanish such that the system remains a FL only at energies, $|\omega|<\omega^*$. 
This suggests that the rare regions of `emerging' local moments, dubbed as Griffiths singularities, act as precursors to a disorder driven QCP that was heretofore unidentified. 
The systematic feedback of the instabilities induced by these local moments, into the 
underlying FL from which they emerge, give rise to this intrinsic energy scale which would manifest as disorder induced nFL excitations in proximity to the QCP.
We speculate that the $\omega^*\rightarrow 0$ signals the onset of a Griffiths phase, since the fraction of sites with $T_K\rightarrow 0$ might act as nucleation centers for clusters of local moments in an otherwise Fermi liquid system, albeit with a distribution of Kondo scales. 

The results presented here provide a first step in understanding the role of spatial fluctuations due to disorder on electron correlations within an efficient computational framework. An essential ingredient missing in these results is the absence of the physics due to the intersite Rudderman-Kittel-Kasuya-Yosida interaction between the emerging local moments. The framework presented here opens an interesting avenue to incorporate 
spatially nonlocal intersite correlations into single particle
quantities in either the charge channel (nearest-neighbor) or the spin channel (exchange) 
in a single theory, including non-local fluctuations due to disorder beyond stat-DMFT. 
This direction is left as a future challenge.

\acknowledgements
SS acknowledges the support from JNCASR, India, where this work was initiated.  
MJ acknowledges support from the NSF Materials Theory grant DMR-1728457 and the NSF EPSCoR Cooperative Agreement No. EPS-1003897 with additional support from the Louisiana Board of Regents.

\appendix
\section{Scattering dynamics for different cluster size}
\label{appendix1}
\begin{figure}[htp!]
\centerline{\includegraphics[clip=,scale=0.4]
                        {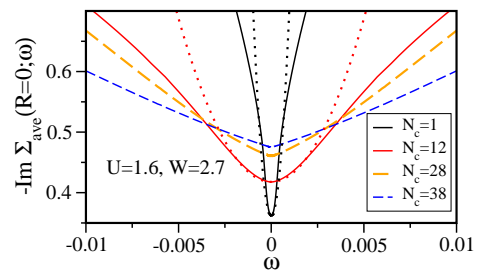}}
  \caption
{{\bf Comparison of the low energy dynamics of the imaginary 
part of the average self-energy for different cluster sizes:}
the imaginary part of the average self-energy, given by 
-Im$\Sigma_{ave}(\mathbf{R}=0,\omega)$ is shown for $U=1.6$ and 
$W=2.7$ on a wider energy range compared to Figure~\ref{fig:Nc_compare}.
The dotted lines represent fits to the Fermi liquid form, given by $\sim a\omega^2$. While small cluster sizes of $N_c=1,\,12$ fit well to this form, the systematic deviation is clearly noticeable for $N_c=28,\,38$.}
  \label{fig:SE_a0_Nc}
\end{figure}

In all our calculations presented in the main text we considered a specific cluster 
size of $N_c=38$. 
In the Figure~\ref{fig:Nc_compare} of main text we demonstrated that the low energy frequency dependence of the disorder averaged self-energy indeed tends to converge to a particular functional form with increasing $N_c$. 
In order to justify this statement we had compared -Im$\Sigma_{ave}(\mathbf{R}=0;\omega)-a_0$ for different $N_c$ at a fixed $W=2.7$. 
In Figure~\ref{fig:SE_a0_Nc} we show the same, but without subtracting the zero frequency component, $a_0$. Similar to the observation for the low energy frequency dependence, the 
$a_0$ values for $N_c=28$ and $N_c=38$ also appear close enough quantitatively.
\begin{figure}[ht!]
\centerline{\includegraphics[clip=,scale=0.16]
                        {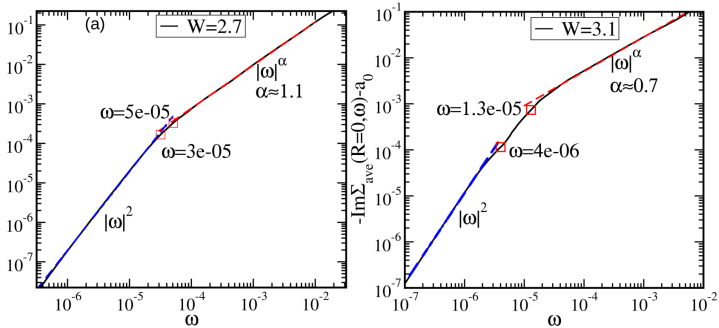}}
  \caption{{\bf Procedure for obtaining the crossover frequency $\omega^*$ and the crossover region:} The disorder averaged electron self-energy, -Im$\Sigma_{ave}(\mathbf{R}=0,\omega)$ is plotted as a function of energy, for disorder strength, (a)$W=2.7$eV and (b)$W=3.1$eV. We notice that there exists a range of frequencies over which the low energy scattering dynamics crosses over from Fermi liquid ($\omega^2$) like to non-Fermi liquid ($|\omega|^\alpha$) behavior. This crossover range, symbolized by $\delta\omega^*$, is bounded by two such frequencies as shown by the red (open) squares in (a) and (b). The respective crossover frequency, $\omega^*$ is thereby estimated as the midpoint of this region. For (a) $W=2.7$eV we obtain $\omega^*\approx 4\times10^{-5}$eV with a crossover range lying approximately between $3\times10^{-5}$eV - $5\times10^{-5}$eV over which the FL excitations crossover to approximately 
  $|\omega|$ like behavior. (b) For $W=3.1$eV we obtain $\omega^*\approx 8\times10^{-6}$eV with a crossover range lying between $4\times10^{-6}$eV - $1.3\times10^{-5}$eV over which the FL excitations crossover to an approximately $|\omega|^{0.7}$ like behavior.}  
  \label{fig:SE_analysis}
\end{figure}
\section{Estimation of the crossover energy scale and range:}
\label{appendix0}
In this section we demonstrate the procedure by which we estimate the $\omega^*$ and the $\delta\omega^*$ plotted in Figure~\ref{fig:nFL_scale} using two representative examples.  The $\delta\omega^*$ obtained from this analysis may be interpreted as a range of frequencies over which the estimated $\omega^*$ may exist.  In Figure~\ref{fig:SE_analysis}(a,b) we plot the disorder averaged electron self-energy, -Im$\Sigma_{ave}(\mathbf{R}=0,\omega)$ (with -Im$\Sigma_{ave}(\mathbf{R}=0,\omega=0=a_0)$ subtracted) as a function of energy for disorder strength, (a)$W=2.7$eV and (b)$W=3.1$eV. We identify a range of frequencies, $\delta\omega^*$, that is bounded by two frequencies $\omega_1$ and $\omega_2$ over which the scattering dynamics crosses over from Fermi liquid ($\omega^2$) like to non-Fermi liquid ($|\omega|^\alpha$) behavior, such that $\omega_1<\omega^*<\omega_2$. The crossover frequency for each disorder strength is thereby estimated as $\omega^*=\frac{\omega_1+\omega_2}{2}$.

\section{Density of states}
\label{appendix2}
\begin{figure}[h!]
\centerline{\includegraphics[clip=,scale=0.25]{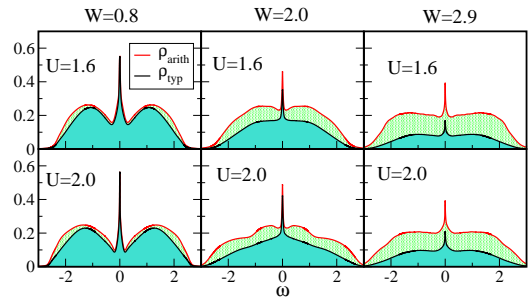}}
\caption{{\bf Density of states:} The arithmetically averaged density of states
$\rho_{arith}(\mathbf{R}=0,\omega)$ (ADOS) is compared with the geometrically
averaged density of states, $\rho_{typ}(\mathbf{R}=0,\omega)$ (TDOS) at two
representative $U=1.6,\, 2.0$ and $N_c=38$, for different $W$'s.}
\label{ADOS2}
\end{figure}

\begin{figure}[h!]
\centerline{\includegraphics[clip=,scale=0.4]{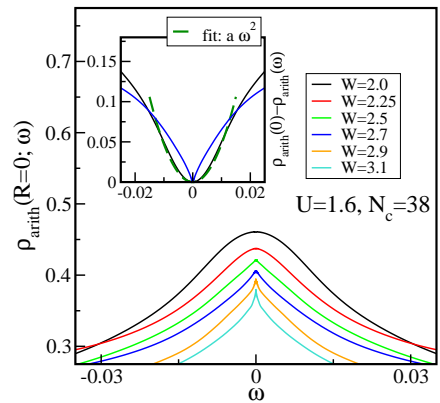}}
\caption{{\bf Density of states at low energy scales: }[Main panel]:
The low energy form of the arithmetically averaged density of states (DoS), $\rho_{arith}(\omega)$
at $U=1.6$ and cluster size, $N_c=38$ is plotted for different disorder strengths, $W$.
The development of cusp in the low energy spectral lineshape is evident.
[Inset]: Two representative
   data for $W=2.0$ and $W=2.7$ are plotted with the
   $\rho_{arith}(\mathbf{R}=0;0)$ subtracted. The low energy
   form of $\rho_{arith}(\mathbf{R}=0;0)-\rho_{arith}(\mathbf{R}=0;\omega)$
   for $W=2.0$ fits well to a form $\sim \omega^2$ as dictated by the
   Fermi liquid (FL) theory where as for $W=2.7$
   clear deviation from the conventional lineshape is evident. However,
   note that at the lowest energy scales a FL form should still hold
   because the self-energy is still a FL at the lowest energy scales.}
\label{ADOS1}
\end{figure}

In this section we analyse the density of states for disorder strengths considering 
$N_c=38$. 
In Fig.~\ref{ADOS2} we compare the average ($\rho_{arith}$) and the typical ($\rho_{typ}$) DoS, for different disorder strengths and two representative parameters for $U$. In agreement with conventional observation for non-interacting disordered systems, $\rho_{typ}$ and $\rho_{arith}$ look similar at low $W$, showing appreciable differences at higher $W$'s. It is worth noting that for $W=2.0$, the Hubbard bands broaden for $U=2.0$ as compared to $U=1.6$, indicating the system at $U=1.6(U/W <1)$ feels an effectively higher interaction strength compared to $U=2.0(U/W =1)$. In the main panel of Fig.~\ref{ADOS1} we demonstrate the evolution of the low energy features of $\rho_{arith}(\mathbf{R}=0,\omega)$ for different $W$'s at $U=1.6$ and $N_c=38$.  

The average DOS (ADOS) also follow similar features as the average self-energy, $\Sigma_{ave}(\mathbf{R}=0;\omega)$ in the sense that with increasing disorder, the ADOS also starts developing a singular (cusp-like) feature on the low energy scales. A clear deviation from the Fermi liquid lineshape can be observed for $W=2.7$ from the inset of Fig.~\ref{ADOS1} where the quantity $\rho_{arith}(\mathbf{R};0)-\rho_{arith}(\mathbf{R};\omega)$ is plotted for $W=2.0$ and $W=2.7$.


\section{Distribution of Kondo scales for lower disorder strengths:} 
\label{appendix3}
\begin{figure}[htp!]
  \centerline{\includegraphics[clip=,scale=0.5]
                        {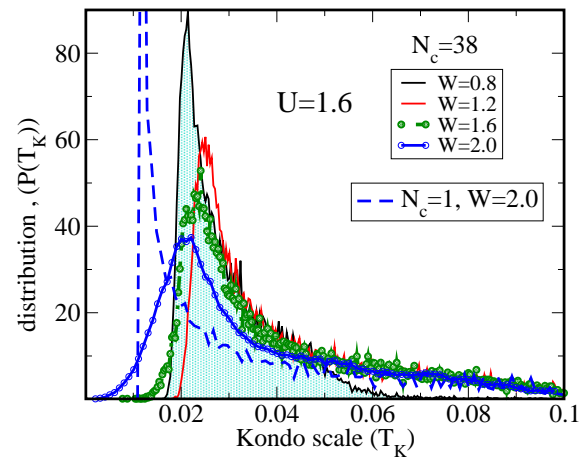}}
\caption
{{\bf Distribution of Kondo scales for lower disorder strengths:} 
The evolution of the distribution of Kondo scales, $P(T_K)$, is demonstrated as a function of increasing the disorder strength, shown for, $W=0.8,\;1.2,\;1.6,\;2.0$ at a fixed interaction strength, $U=1.6$. A well formed peak at an energy, $T_K^{peak}$ can be identified that initially shifts towards higher energy scales and only beyond a certain disorder strength shifts towards lower energy scales. At $W=0.8$, a relatively narrow distribution is obtained in contrast to higher $W$'s where $P(T_K)$ starts developing broad tails on the higher $T_K$ side and also tails reaching lower and lower $T_K$'s as $W$ is increased. The $N_c=1$ limit is also shown as a blue dashed curve.}
\label{fig:low_disorder_TK}
\end{figure}

In Section~\ref{PTk} we only discussed the distribution of Kondo scales for  
moderate to high disorder values. In the following we briefly discuss the 
respective $P(T_K)$ obtained for lower disorder strengths and also compare 
them with the $N_c=1$ or DMFT limit.
In Fig.~\ref{fig:low_disorder_TK}, we plot $P(T_K)$ for several $W$'s ranging from $W=0.8$ to $W=2.0$.  A well defined peak at an energy scale, $T_K^{peak}$ can be identified for these disorder strengths. Such $T_K^{peak}$'s were also identified within our TMT-DMFT calculations, and were identified as universal low 
energy scales, within a local theory (see \cite{my_paper_TMT} for details). In concurrence with the local theory, $T_K^{peak}$ initially increases and only beyond a certain $W$ does it start decreasing, reflecting upon an initial disorder-screening of $U$ followed by a subsequent co-operative effect where both $W$ and $U$ tend to suppress the effective hybridization resulting in reduced charge fluctuations and thus manifesting as a reduced Kondo scale \cite{my_paper_TMT}. However, unlike a local theory, inclusion of short-range correlation effects of disorder, leads to the emergence of a low-$T_K$ tail \cite{Vlad_griffiths_phase} that was completely absent in the TMT-DMFT calculations. In Fig.~\ref{fig:low_disorder_TK} this fact is illustrated as blue-dashed line for $N_c=1$ and as blue-solid line with open circles for $N_c=38$ and for a particular disorder strength of $W=2.0$. For a relatively low disorder, $W=0.8$, the distribution is narrower, while long tails spanning a wider range of $T_K$'s develop as $W$ is gradually increased. More importantly, systematic inclusion of short range correlation effects makes the system explore low energy scales that were left untrod within a local theory.
\newpage
\bibliography{ref}
\end{document}